\documentclass[pre,aps,showpacs,12pt,reprint,superscriptaddress]{revtex4-1}

\usepackage[utf8]{inputenc}  
\usepackage[T1]{fontenc}
\usepackage{lmodern}         

\usepackage{amsmath, amssymb, amsfonts}
\usepackage{mathtools}
\usepackage{bm}
\usepackage{mathrsfs}
\usepackage{yfonts,bbm}
\usepackage{slashed}

\usepackage{graphicx}
\usepackage{tikz}
\usepackage{adjustbox}
\usepackage{booktabs}
\usepackage{array}
\usepackage{multirow}

\usepackage{chemarrow}

\usepackage[hidelinks]{hyperref}
\hypersetup{
  colorlinks=true,
  citecolor=blue,
  urlcolor=blue,
  linkcolor=blue
}

\usepackage{notes2bib}
\bibnotesetup{note-name =,use-sort-key = false}

\DeclareMathAlphabet{\mathpzc}{OT1}{pzc}{m}{it}

\begin{document}

\title{Thermodynamic Circuits: Modeling chemical reaction networks with nonequilibrium conductance matrices}

\author{Paul Raux}
\affiliation{Université Paris-Saclay, CNRS/IN2P3, IJCLab, 91405 Orsay, France }
\affiliation{Université Paris Cité, CNRS, LIED, F-75013 Paris, France}
\author{Christophe Goupil}
\affiliation{Université Paris Cité, CNRS, LIED, F-75013 Paris, France}
\author{Gatien Verley}
\affiliation{Université Paris-Saclay, CNRS/IN2P3, IJCLab, 91405 Orsay, France }

\date{\today}

\begin{abstract}
We derive the nonequilibrium conductance matrix for open stationary Chemical Reaction Networks (CRNs) described by a deterministic mass action kinetic equation. As an illustration, we determine the nonequilibrium conductance matrix of a CRN made of two pseudo-linear sub-networks, called chemical modules, in two different ways: First by computing the nonequilibrium conductances of the modules that are then serially connected. Second by computing the nonequilibrium conductance of the CRN directly. The two approaches coincide, as expected from our theory of thermodynamic circuits. 
\end{abstract}

\maketitle

\section{Introduction}
Open Chemical Reaction networks (CRNs) are paradigmatic examples of complex out of equilibrium systems. Over the last decades, they have attracted a long standing attention as they combine the theory of graphs and hyper-graphs \cite{Schnakenberg1976, Hill1989, DalCengio2023}, dynamical systems theory \cite{Ross2008, vanKampen2007} and thermodynamics \cite{Nicolis1977, Schmiedl2007, Rao2016}. CRNs are of various complexity, from pseudo-linear dynamics highly similar to Markov jump processes \cite{Avanzini2024}, to non-linear dynamics with interacting species (beyond interaction through chemical reactions only) \cite{Avanzini2020}, passing by complex balance dynamics for deficiency zero CRNs~\cite{Anderson2015}. Given their complexity, the circuit decomposition of CRNs is appealing to simplify the study of each chemical module separately. Once characterized, each chemical module can be reused in other CRNs without further studies while the global investigation of a CRN require to restart from scratch upon any minor modification of the network. An existing circuits approach relies on chemical modules modeled by current-concentration characteristics~\cite{Avanzini2023}. In the latter work, constraints on internal species are dealt indirectly with the help of emergent cycles~\cite{Polettini2014}, i.e., a set of reactions that upon completion do not change the internal concentrations, but only transfer matter between chemostats. Moreover, couplings between chemical currents are not at the core of this theory although they significantly constrain the efficiency of chemical transduction~\cite{Caplan1966, Wachtel2022, Bilancioni2025, Vroylandt2018}. 

In the present work, we extend to circuits of chemical modules our thermodynamic circuit theory~\cite{Raux2024}. The description of chemical modules by nonequilibrium conductance matrices combines the simplification of circuit decomposition with the ability to lift the notion of coupling between reaction currents to the level of matter currents exchanged with the chemostats.  Additionally, nonequilibrium resistance/conductance matrices provide a model for chemical transduction within irreversible thermodynamics that accounts for chemical currents coupling. Therefore, energy conversion (e.g., thermoelectric \cite{Raux2025thermoelec}) and chemical transduction are essentially studied on the same footing, illustrating the universality of concepts in the physics of conversion processes.

The paper is organized as follows: 
In section \ref{sec : chemical noneq cond matrix}, we derive the nonequilibrium resistance/conductance matrices at various levels of description of the open CRN. We start by recalling the theory of chemical kinetics emphasizing its thermodynamic consistency~\cite{ Polettini2014}. Then, given their prominence in our work, we determine the conservation laws relating the chemostat currents, i.e., the physical currents corresponding to matter received by the CRN from the chemostats. This approach transfers Schnakenberg's decomposition of reaction currents on cycle currents~\cite{Schnakenberg1976} to higher level where physical currents are decomposed on fundamental currents~\cite{Raux2024}. This allows for an effective description of the stationary CRN that does not rely on emergent cycles~\cite{Polettini2014, Avanzini2023}. Instead, we consider conservation laws of physical currents (i.e. from chemostats). This simple change of viewpoint makes easier the connection with our theory of thermodynamic circuits and has greater similarity with the framework of Markov jump processes.
In section \ref{illustration}, we illustrate the calculation of the chemical nonequilibrium conductance matrix on the two first chemical modules appearing in Ref.~\cite{Avanzini2023}. Then, we provide two derivations of the nonequilibrium conductance matrix of the CRN built upon their serial association.
First, we apply our law of resistance addition for serial association of thermodynamic devices. Second, we derive the nonequilibrium conductance matrix directly for the whole network. 


%
\section{Chemical nonequilibrium conductance matrix}
\label{sec : chemical noneq cond matrix}
In this section, we fix the notation by recalling the stochastic thermodynamics of CRNs \cite{Polettini2014}. Whenever possible, we make connection with stationary Markov jump processes as studied in the first section of Ref.~\cite{Raux2024}. In the same spirit, we review the description of conservation laws and conserved quantities existing in closed and opened CRNs. Finally, we build the nonequilibrium conductance matrix describing the current--force characteristics of a CRN. 
\subsection{Chemical kinetics}
We describe a chemical reaction network by a set of chemical species $Z_{\alpha}$ of concentration $[Z_\alpha]$, identified by $\alpha \in \mathscr{S}=\lbrace 1,2,...,|\mathscr{S}| \rbrace$ that are transformed by chemical reactions denoted by the index $\rho \in \mathscr{R}=\lbrace 1,2,...,|\mathscr{R}| \rbrace$. Each reaction $\rho$ is arbitrarily oriented, is assumed to be reversible and follows a chemical equation of the form:
\begin{equation}
\sum_{\alpha\in\mathscr{S}}\bm \nabla_{\alpha,\rho}^+  Z_\alpha{}
\; \underset{k^{-}_{\rho}}{\overset{k^{+}_{\rho}}{\rightleftharpoons}} \;
\sum_{\alpha\in\mathscr{S}}\bm \nabla_{\alpha,\rho}^- Z_\alpha
\end{equation} 
where $\nabla_{\alpha,\rho}^+$ (respectively $\nabla_{\alpha,\rho}^-$) is the number of molecules $\alpha$ consumed (respectively produced) for a given forward reaction $\rho$ \cite{Rao2016}.
The evolution of the concentrations of species $\alpha$ follows the kinetic equation
\begin{equation}
\frac{d[\bm Z]}{dt}=\bm \nabla \bm j + \bm{\mathcal{I}}
\label{eq : rate equation}
\end{equation}
where $[\bm Z]$ is the concentration vector of components $[Z_\alpha]$, $\bm \nabla$ the stoichiometric matrix of components $\nabla_{\alpha,\rho}=\nabla_{\alpha,\rho}^- -\nabla_{\alpha,\rho}^+$, $\bm j$ the vector of reaction currents and $\bm{\mathcal{ I}}$ the currents exchanged with the chemostats. 
We remark that Eq.~\eqref{eq : rate equation} is a continuity equation for the species concentrations analogous to the master equation describing Markov processes. In this analogy, the species concentrations correspond to state probabilities, the stoichiometric matrix $\bm \nabla$ to the incidence matrix of the graph of the Markov jump process, and the reaction currents $\bm j$ to the edge probability currents. The source term $\bm{\mathcal{I}}$ appears in the kinetic equation, but not in the master equation. This term is required to have fixed concentration of the external species. There is no such term for Markov jump processes: the reservoirs constrains the transition rates and not the state occupancy directly. Another difference is that the master equation ruling the time evolution of the state probability for Markov jump processes is linear, while the kinetic equation ruling the time evolution of the concentrations can be non linear.

Assuming mass action law, the reaction fluxes read 
\begin{equation}
j_\rho \equiv k_\rho^+[\bm Z]^{\bm \nabla_\rho^+} - k_\rho^-[\bm Z]^{\bm \nabla_\rho^-}
\label{eq : def reaction currents}
\end{equation}
where we denote $k_\rho^\pm$ the kinetic rates and $\bm \nabla^\pm_\rho$ the $\rho$'s  column of matrix $\bm \nabla^\pm$. 
We use the notation $\bm x^{\bm y}=\prod_\alpha x_\alpha^{y_\alpha}$. The kinetic rates are chosen according to the local detailed balance  \cite{Avanzini2024}:
\begin{equation}
RT\log \frac{k_\rho^+}{k_\rho^-}=- (\bm \nabla^T \bm \mu^0(T) )_\rho
\label{eq : local detailed balance}
\end{equation}
where $\bm \nabla^T$ is the transpose of the stoechiometric matrix, $\bm \mu^0$ the column vector of standard chemical potential for species in $\mathscr{S}$, $T$ the temperature of the isothermal CRN and $R$ the perfect gaz constant. With words, local detailed balance relates dynamics to thermodynamics,  i.e. kinetic rates to standard chemical potentials. It does so in order to ensure dynamically consistent equilibrium or stationary nonequilibrium states. 
We define the force conjugated to the reaction currents $j_\rho$ as
\begin{equation}
f_\rho \equiv RT\log \frac{k_\rho^+[\bm Z]^{\bm \nabla_\rho^+}}{k_\rho^- [\bm Z]^{\bm \nabla_\rho^-}}=-(\bm \nabla^T \bm \mu )_\rho=-\Delta_\rho G
\label{eq : reaction affinity}
\end{equation}
where the vector $\bm \mu$ gathers the chemical potential of the species in $\mathscr{S}$ and $\Delta_\rho G$ is the Gibbs free energy change caused by reaction $\rho$. The component $\alpha$ of the chemical potential vector $\bm \mu$ reads \cite{Avanzini2024}
\begin{equation}
\mu_\alpha=\mu_{\alpha}^0+RT\log[Z_\alpha].
\end{equation}
By definition, chemostats set to constant values the concentrations of external species denoted $[Y_\alpha]$, with $\alpha \in \mathscr{S}_\textsc{y}$ the subset of chemostated species. The concentrations of internal species denoted $[X_\alpha]$ are free, i.e., species with $\alpha \in \mathscr{S}_\textsc{x}$ are not exchanged with any chemostat. The total set of species is the disjoint union of these two sets
\begin{equation}
\mathscr{S}=\mathscr{S}_\textsc{x}\cup\mathscr{S}_\textsc{y}.
\end{equation}
Accordingly, the concentration vector and the stoichiometric matrix writes respectively
\begin{equation}
	[\bm Z] = \begin{pmatrix} [\bm X] \\ [\bm Y] \end{pmatrix}, \qquad
\bm \nabla =
\begin{bmatrix}
\bm \nabla_\textsc{x} \\
\bm \nabla_\textsc{y}
\end{bmatrix}.
\end{equation}
The rate equation rewrites as
\begin{equation}
\displaystyle\frac{d[\bm Z]}{dt}
=\begin{pmatrix}
\displaystyle \frac{d[\bm X]}{dt}\\
\bm 0
\end{pmatrix}
=\begin{bmatrix}
\bm \nabla_\textsc{x}\\
\bm \nabla_\textsc{y}
\end{bmatrix} \bm j + \begin{pmatrix}
\bm 0 \\
\bm i
\end{pmatrix},
\label{eq : rate equation split}
\end{equation}
where we use $\frac{d [\bm Y]}{dt}=\bm 0$ for external species. Hence, the splitting between external and internal species of Eq.~\eqref{eq : rate equation split} leads to a definition of the currents received by the open CRN from the chemostats
\begin{equation}
\bm i=-\bm \nabla_\textsc{y} \bm j,
\label{eq : definition des courants de chemostats}
\end{equation}
and a kinetic equation for internal species with no source term
\begin{equation}
\frac{d[\bm X]}{dt}=\bm \nabla_\textsc{x} \bm j
\label{eq : internal species evolution}
\end{equation}
since internal species are not exchanged with the environment, i.e. $\bm{\mathcal{I}}=(\bm 0,\bm i)^T$ by definition. In this work, $\bm i$ is the vector of chemostat currents as its components are matter currents only, as compared to Ref.~\cite{Raux2024} in which they are physical currents of any extensive quantity exchanged with a reservoir. 

%
\subsection{Conservation laws}
Let's determine the conservation laws relating physical currents. We assume that the stoichiometric matrix $\bm \nabla$ has a non zero cokernel, i.e., it exists $\bm L$ such that
\begin{equation}
\bm L \bm \nabla=0.
\label{eq : definition conservation laws}
\end{equation}
Then, each row of matrix $\bm L$ is a left eigenvector of the stoichiometric matrix with null eigenvalue. We denote by $|\mathscr{L}_\mathrm{cl}|$ the number of rows of matrix $\bm L$, label them with $\lambda \in \{1,2,\dots, |\mathscr{L}_\mathrm{cl}| \}$. The subscript $\mathrm{cl}$ stands for "close" as will become clear later on. Since we consider open CRNs, we can split $\bm L$ block wise by columns (splitting between the internal and external species)
\begin{equation}
\bm L =
\begin{bmatrix}
\bm \ell_\textsc{x} &  \bm \ell_\textsc{y}
\end{bmatrix},
\label{eq : X Y splitting of the conservation laws}
\end{equation}
as the first lines of $\bm \nabla$ are for the internal species, and the final ones for external species.
Multiplying the rate equation Eq.~\eqref{eq : rate equation split} by $\bm L$ yields:
\begin{equation}
\frac{d\bm M}{dt}=\bm \ell_\textsc{y} \bm i,
\label{eq : evolution des quantite conservees}
\end{equation}
where we have defined the moiety vector 
\begin{equation}
\bm M \equiv \bm \ell_\textsc{x} [\bm X].
\end{equation}
A way to interpret Eq.~\eqref{eq : evolution des quantite conservees} is to consider the closed CRN case. Indeed, in this case $\bm i=\bm 0$ and Eq.~\eqref{eq : evolution des quantite conservees} reveals that $\bm M$ gathers the conserved quantities of the dynamics. $\bm L$ is thus the matrix whose lines are the conservation laws of the CRN if it was closed. This explains why $\bm L$ is usually called the matrix of conservation laws in the literature. For open CRNs, a conserved quantity $ M_\lambda$ remains conserved after opening the network if $(\bm \ell_\textsc{y} \bm i)_\lambda = 0$. But, for an open CRN, $\bm i\neq \bm 0$ by definition. Then, $M_\lambda$ is a conserved quantity of the dynamics only if $ \forall \alpha \in \mathscr{S}_\textsc{y}$, $(\bm \ell_\textsc{y})_{\lambda \alpha}= \bm 0$ (case u) or if $(\bm \ell_\textsc{y} \bm i)_\lambda=0$ with $\lambda$th line of $\bm \ell_\textsc{y}$ having at least one non zero component. Note that without any assumptions on the chemostat currents, $(\bm \ell_\textsc{y} \bm i)_\lambda\neq 0$ (case b). The corresponding moeities are thus in general no longer conserved quantities of the dynamics.  
According to these two cases, we split linewise the matrix $\bm L$ and the vector $\bm M$ as
\begin{equation}
\bm L =
\begin{bmatrix}
\bm \ell^u \\
\bm \ell^b
\end{bmatrix}, \quad \bm M 
=\begin{bmatrix}
\bm M^u\\
\bm M^b
\end{bmatrix}.
\label{eq : U B splitting for the conservation laws and conserved quantities}
\end{equation}
The matrix $\bm \ell^u$ gathers the $|\mathscr{L}^{u}|$ conservation laws such that the components of $\bm M^u$ remain conserved whatever the current incoming from the reservoir: Opening the CRN always preserves the conservation laws $\bm \ell^u$ that are said "unbroken". By definition $\bm \ell_\textsc{y}^u=\bm 0$ and the unbroken conservation laws take the form 
\begin{equation}
\bm \ell^u=\begin{bmatrix}
\bm \ell_\textsc{x}^u & \bm \ell_\textsc{y}^u
\end{bmatrix} = \begin{bmatrix}
\bm \ell_\textsc{x}^u & 0
\end{bmatrix}. \label{eq : unbroken conservation laws}
\end{equation}
 On the contrary, $\bm \ell^b$ gathers the $|\mathscr{L}^{b}|$ conservation laws that can be associated to conserved quantities only when taking into account the matter exchanged with the chemostats. In this case, a particular combination of the currents incoming from the reservoirs is required to get a constant moiety. If not, the moiety is not conserved and opening the CRN has ``broken'' the conservation law. Applying now the broken conservation laws $\bm \ell^b=\begin{bmatrix}
 \bm \ell_\textsc{x}^b &  \bm \ell_\textsc{y}^b
 \end{bmatrix} = \begin{bmatrix}
 \bm \ell_\textsc{x}^b & \bm \ell
 \end{bmatrix}$ on Eq.~\eqref{eq : rate equation split}, 
 we obtain
 \begin{equation}
     \frac{d \bm M^{b} }{dt} = \bm \ell \bm i
 \end{equation}
%
We shorten the notation of $\bm \ell^b_\textsc{y} $ by $ \bm \ell$, for consistency with Ref.~\cite{Raux2024}. Indeed, when assuming that any species has reached its stationary concentration (as done in the following sections)
\begin{equation}
\bm \ell \bm i= \bm 0.
\label{eq : lois de conservations chemostats}
\end{equation}
Then, $\bm \ell$ stands for the conservation law matrix for physical currents flowing from the reservoirs. In the case of CRNs, one can check that the $|\mathscr{L}| = |\mathscr{L}^{b}| =|\mathscr{L}_{\mathrm{cl}}| - |\mathscr{L}^{u}|$ rows of matrix $\bm \ell$ are indeed the conservation laws that relate linearly the chemostat currents at stationary state. Hence, we have exhibited in this section the block form of matrix 
\begin{equation}
	\bm L = 
	\begin{bmatrix}
		\bm \ell_\textsc{x}^u & 0 \\
		\bm \ell_\textsc{x}^b & \bm \ell
	\end{bmatrix}, \label{Cannonical-Cons-Laws}
\end{equation}
in which appears the submatrix $\bm \ell$ of conservation laws for chemostat currents at the core of our theory of thermodynamic circuits. We provide in the next section a direct way of getting the conservation law matrix $\bm{\ell}$ from the stoechimoetric and cycle matrices.

\subsection{Cycles and selection matrix}
\subsubsection{Cycles}
\label{subsub : cycles}
From now on, we assume that the open CRN has reached a nonequilibrium stationary state. Then, Eq.~\eqref{eq : internal species evolution} reduces to 
\begin{equation}
\bm \nabla_\textsc{x} \bm j = \bm 0
\label{eq : redundant reaction currents}
\end{equation}
and the reaction currents are linearly dependent: the lines of $\bm \nabla_\textsc{x}$ contains the coefficients of a vanishing linear combination of reaction currents. Hence, it exists sequences of reactions, called cycles, that let the concentrations of internal species unchanged. We thus have 
\begin{equation}
\bm j=\bm {\mathcal{C}}\bm J,
\label{eq : def cycle currents}
\end{equation}
where $\bm {\mathcal{C}}$ is the cycle matrix whose $|\mathscr{C}|$ columns are basis vectors of $\mathrm{ker}(\bm \nabla_\textsc{x})$ such that $\bm \nabla_\textsc{x} \bm j = \bm \nabla_\textsc{x} \bm {\mathcal{C}} \bm J = 0$. We denote as usual the cycle currents $\bm J$: it is the vector of independent currents among the reaction currents in $\bm j$. As already emphasized in Ref.~\cite{Raux2024, Raux2025thermoelec}, $\bm {\mathcal{C}}$ is also analogous to a selection matrices. 

Now that we have the linearly independent cycle currents at hand, we can provide a direct computation of the matrix $\bm {\ell}$ for the conservation laws constraining chemostat currents. Indeed, Eqs. (\ref{eq : definition des courants de chemostats}, \ref{eq : lois de conservations chemostats}, \ref{eq : def cycle currents}) lead to 
\begin{equation}
	\bm \ell \bm i =-\bm \ell  \bm \nabla_\textsc{y} \bm j =-\bm \ell  \bm \nabla_\textsc{y} \bm {\mathcal{C}} \bm J  = \bm 0.
\end{equation}
Since the currents $\bm J$ are linearly independent, the above equation imposes 
\begin{equation}
	\bm \ell \bm \phi = 0 \quad \text{with } \bm \phi \equiv \bm \nabla_\textsc{y} \bm {\mathcal{C}}
\end{equation}
the matrix of physical exchanges whose columns provide the matter exchanged with chemostats during each cycle. In words, the rows of matrix $\bm \ell $ are the row vectors of current conservation laws. They belong to the cokernel of $ \bm \phi $. These row vectors indicate the exact proportion of chemostat currents that when imposed to the system does not change the internal species concentrations. Our approach of thermodynamic circuits based on current conservation laws is an alternative to the method relying on emergent cycles~\cite{Avanzini2023,Polettini2014}. The vectors associated to such emergent cycles belong to the kernel of the stochiometric sub-matrix for internal species, but are not in the kernel of the stochiometric sub-matrix for external species. Said differently, emergent cycle are vectors (in the space of reactions) that indicate a set of reactions that upon completion do not change the internal concentrations but do change the external ones. In the end, conservation laws for chemostat currents and emergent cycles are two different ways of keeping internal species fixed while transfering species between chemostats : the former focuses on the chemostat currents and the latter on cycles of reactions that only transfer species from one chemostat to another.


\subsubsection{Selection matrix}
\label{subsub : selection matrix}
Like reaction currents are linear combination of independent cycle currents, we obtain below physical (chemostat) currents as linear combinations of the fondamental currents. 
The conservation laws write $\bm \ell \bm i=0$ and the physical currents are linearly dependent as expected. We can select a subset of linearly independent currents as 
\begin{equation}
\bm i=\bm S \bm I
\label{eq : fundamental chemostats currents}
\end{equation}
where $\bm S$ is the selection matrix whose columns are basis vector of $\mathrm{ker}(\bm \ell)$. 
We denote as usual $\bm I$ the fundamental currents: it is the vector of independent currents among the physical currents in $\bm i$. By definition, the matrix $\bm S$ has linearly independent columns and admits a left pseudo inverse $\bm S^{+}$. The fundamental current vector then write $\bm I = \bm S^{+} \bm i$. We emphasize that the fundamental currents are chosen as linearly independent currents among the chemostat currents. Depending on the CRN, we might have identical cycle and fundamental currents $\bm J=\bm I$, in which case $\bm S = -\bm \nabla_\textsc{y} \bm {\mathcal{C}}$. However, this is not generally the case as we show in appendix \ref{fund or cycle forces}.

\subsection{Entropy production and thermodynamic forces}
In this section, we determine the thermodynamic forces conjugated to reaction, cycle, chemostat and fundamental currents for a stationary open CRN. This is done by ensuring thermodynamic consistency, i.e., identical Entropy Production Rate (EPR) at all levels of description of the CRN~\cite{Schnakenberg1976}. We start from the EPR $\sigma$ at the level of reaction forces and currents given (up to the temperature factor) by
\begin{equation}
T\sigma=\bm f^T  \bm j  \qquad \text{(reaction level)}.
\label{eq : EPR (1)}
\end{equation}
We switch to the cycle level by inserting the definition of the cycle currents Eq.~\eqref{eq : def cycle currents} in the EPR Eq.~\eqref{eq : EPR (1)}. The EPR then reads 
\begin{equation}
T\sigma=\bm F^T  \bm J  \qquad \text{(cycle level)}
\end{equation}
where we have introduced the cycle affinity
\begin{equation}
\bm F^T =\bm f^T\bm {\mathcal{C}}.
\label{eq : definition of the cycle affinity}
\end{equation}
Another decomposition of the EPR follows from inserting the definition of the reaction affinity Eq.~\eqref{eq : reaction affinity} in Eq.~\eqref{eq : EPR (1)}. Using in addition that the rate equation in stationary state yields $\bm{\mathcal{I}}=-\bm \nabla \bm j$ , we obtain
\begin{equation}
T\sigma=\bm \mu^T \bm{\mathcal{I}}
\label{eq : mu/I EPR}
\end{equation}
in agreement with $\bm f^T = - \bm \mu^T \bm \nabla$ of Eq.~\eqref{eq : reaction affinity}.
This EPR further simplifies by noting that $\bm{\mathcal{I}}^T=\begin{pmatrix}
\bm 0&\bm i
\end{pmatrix}$ and by using the internal/external splitting of $\bm \mu^T = \begin{pmatrix}
\bm \mu_\textsc{x}^T & \bm a^T
\end{pmatrix}$ as 
\begin{equation}
T\sigma=\bm a^T \bm i  \qquad \text{(physical level)}.
\end{equation}
where $\bm a = \bm \mu_\textsc{y}$ is the vector of chemical potentials for external species.
We can finally use the redundancy of the chemostat currents 
to write the EPR as
\begin{equation}
T\sigma=\bm A^T  \bm I  \qquad \text{(fundamental level)}
\label{eq : EPR (4)}
\end{equation}
where we have identified the fundamental force vector 
\begin{equation}
\bm A=\bm S^T \bm a
\label{eq : fundamental affinities}
\end{equation}
conjugated to the fundamental current vector $\bm I$. This ends the identification of the four relevant current-force decompositions preserving the EPR.

\subsection{Non equilibrium conductance matrix}
We now turn to the determination of the current-force characteristics based at each level of description on a nonequilibrium conductance matrix. 
We first define the reaction resistance matrix by the diagonal matrix 
\begin{equation}
\bm r=\mathrm{diag}(r_1,\dots, r_{|\mathscr{R}|}) \text{ with } r_\rho=\frac{f_\rho}{j_\rho}.
\label{eq : reaction resistance matrix}
\end{equation}
The thermodynamic force at the level of reactions writes as a function of reaction currents
\begin{equation}
\bm f = \bm r \bm j.
\end{equation}
Then, the cycle currents and forces are related by
\begin{equation}
\bm F = \bm R \bm J, \text{ with }\bm R\equiv \bm {\mathcal{C}}^T \bm r \bm {\mathcal{C}} \qquad \text{(cycle level)}
\label{eq : cycle resistance matrix}
\end{equation}
since
\begin{equation}
\bm F=\bm {\mathcal{C}}^T \bm f=\bm {\mathcal{C}}^T\bm r \bm j=\left( \bm {\mathcal{C}}^T \bm r \bm {\mathcal{C}} \right) \bm J
\label{eq : affinity-flux cycle relation}
\end{equation}
where we use Eqs.~(\ref{eq : def cycle currents},\ref{eq : definition of the cycle affinity}). 
The matrix $\bm R$ is symmetric and positive semi definite by non-negativity of the EPR. Assuming an inverse matrix exists, we call $\bm R^{-1}$ the cycle conductance matrix. It relates the cycle currents to its conjugated affinities as
\begin{equation}
\bm J = \bm R^{-1}\bm F.
\label{eq : cycle conductance matrix}
\end{equation} 

Another conductance matrix exists relating stationary currents $\bm {\mathcal{I}}$ and chemical potential via
\begin{equation}
\bm{\mathcal{I}}=\bm{\mathcal{G}}\bm \mu, \text{ with }
\bm{\mathcal{G}}\equiv \bm  \nabla \bm r^{-1} \bm \nabla^T,
\end{equation}
since
\begin{equation}
\bm{\mathcal{I}}=-\bm \nabla \bm j = -\bm \nabla \bm r^{-1} \bm f = (\bm \nabla \bm r^{-1} \bm \nabla^T) \bm \mu .
\end{equation}

In a similar way, we obtain the conductance matrices at the level of physical forces and currents
\begin{eqnarray}
\bm{i}&=&\bm{g}\bm a, \text{ with }
\bm g \equiv \bm {\phi} \bm R^{-1} \bm {\phi}^T \label{eq : chemostat conductance matrix}. \\ && \text{(physical level)} \nonumber 
\end{eqnarray}
To do so, we start from Eq.~\eqref{eq : definition des courants de chemostats} and use Eqs.~(\ref{eq : def cycle currents},\ref{eq : cycle conductance matrix},\ref{eq : definition of the cycle affinity}) to get
\begin{equation}
\bm i= - \bm \nabla_\textsc{y} \bm {\mathcal{C}} \bm R^{-1} \bm {\mathcal{C}}^T \bm f.
\label{eq : chemostat conductance matrix 2}
\end{equation}
The last step involves the reaction affinity of Eq.~\eqref{eq : reaction affinity} in vector notation
\begin{equation}
\bm f = - \bm \nabla_\textsc{x}^T \bm \mu_\textsc{x} - \bm \nabla_\textsc{y}^T \bm a  \quad \Rightarrow \quad \bm {\mathcal{C}}^T \bm f = - \bm {\mathcal{C}}^T \bm \nabla_\textsc{y}^T \bm a 
\end{equation}
since $\bm \nabla_\textsc{x} \bm {\mathcal{C}} = 0 $ by definition. We notice that the nonequilibrium conducance matrix at the physical level takes the same expression as for Markov jump processes \cite{Raux2024, Vroylandt2018} once the relevant physical matrix $\bm \phi 
$ has been identified.
We remark that the physical conductance matrix writes explicitly $\bm g = \bm \nabla_\textsc{y} \bm {\mathcal{C}} \bm {\mathcal{C}}^{+}\bm r^{-1}  \left(\bm {\mathcal{C}} \bm {\mathcal{C}}^{+}\right)^{T}  \bm \nabla_\textsc{y}^T$, in which $\bm {\mathcal{C}} \bm {\mathcal{C}}^{+}$ is a projector and not the identity since $ \bm {\mathcal{C}}^{+}$ is a left inverse only.

Finally, the conductance for fundamental currents and forces writes
\begin{equation}
\bm I = \bm G \bm A, \text{ with } \bm G \equiv \bm S^{+} \bm g \bm S^{T+} \quad \text{(fundamental level).}
\label{eq : fundamental conductance matrix}
\end{equation}
Indeed, assuming that $\bm G$ exists such that $\bm I = \bm G \bm A$ and using Eqs.~(\ref{eq : fundamental chemostats currents},\ref{eq : fundamental affinities})  yields
\begin{equation}
    \bm i = \bm S \bm I = \bm S \bm G \bm S^T \bm a, \quad \Rightarrow \quad \bm g = \bm S \bm G \bm S^T.
\end{equation}
that can be inverted to get the fundamental conductance matrix of Eq.~\eqref{eq : fundamental conductance matrix} since $\bm S^+$ is a left inverse. This concludes our derivation of the nonequilibrium conductance matrix associated to a CRN at any level of description. In the end, an analogy with Markov jump processes comes out clearly thanks to the identification of the matrix of physical exchanges $\bm \phi $ on one side and, on another side, considering that the cycle matrix $\bm {\mathcal{C}}$ must be in the kernel of $\bm \nabla_\textsc{x} $ the stochiometric submatrix for internal species only.
\section{Illustration}
\label{illustration}
\begin{figure}
\centering 
\includegraphics[width=0.7\columnwidth]{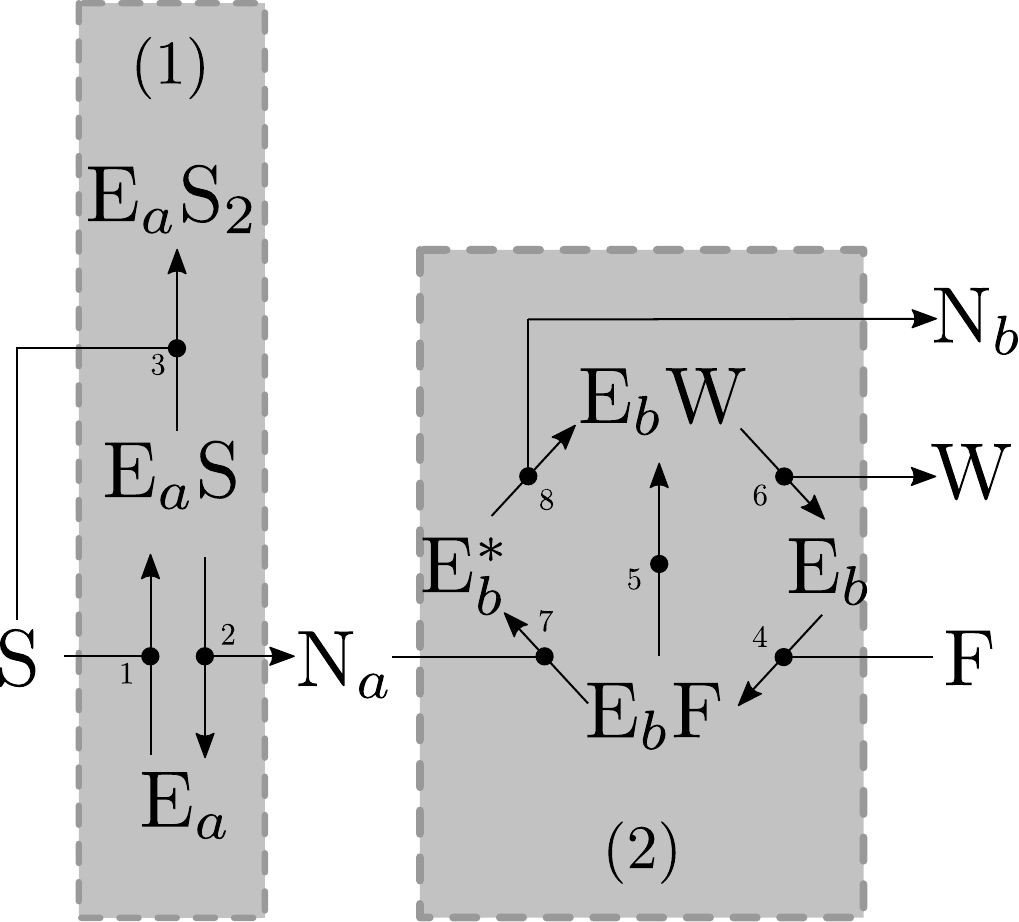}
\caption{CRN decomposed into chemical modules 1 and 2 as proposed in Ref.~\cite{Avanzini2023}. The modules are associated in series. Module 1 has 2 external species $\mathrm{S},\mathrm{N}_a$ and 3 internal species $\mathrm{E}_a,\mathrm{E}_a\mathrm{S},\mathrm{E}_a\mathrm{S}_2$. Module 2 has 4 external species $\mathrm{N}_a,\mathrm{F},\mathrm{W},\mathrm{N}_b$ and 4 internal species $\mathrm{E}_b,\mathrm{E}_b\mathrm{F},\mathrm{E}_b^*,\mathrm{E}_b\mathrm{W}$. The serial association is implemented by ensuring the conservation of the reaction currents $j_{2}=j_{7}$ and the equality of the stationary concentration of $\mathrm{N}_a$ computed in modules 1 and 2. \label{fig : illustration }}
\centering
\includegraphics[width=0.7\columnwidth]{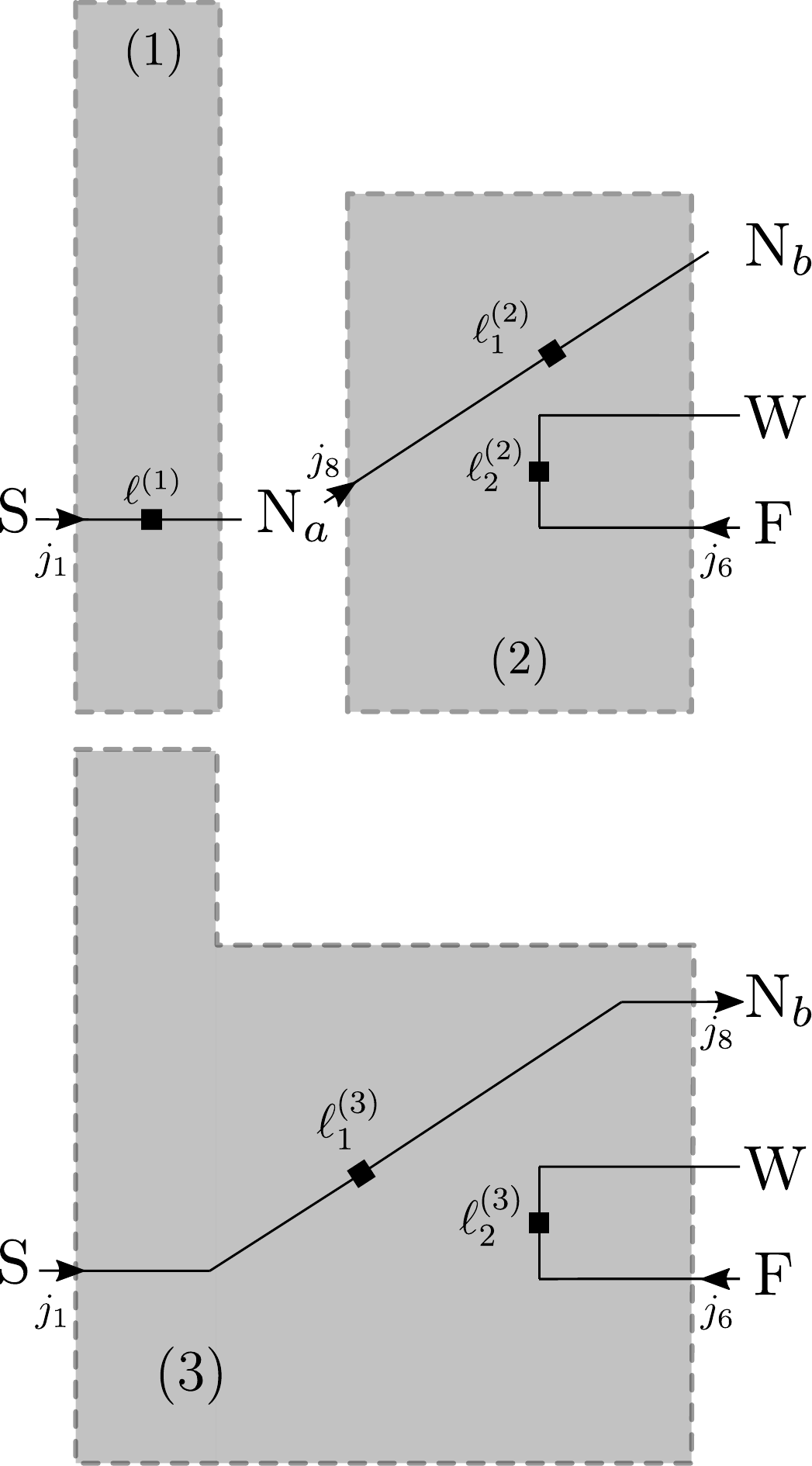} 
\caption{Effective description of modules 1, 2 and 3. In modules 2 and 3, a coupling exists between currents $j_6$ and $j_8$, although it does not appear on this graphical representation of conservation laws. \label{fig : effective modules}}
\end{figure}

We now derive the non equilibrium conductance matrix for the CRNs of Fig.~\ref{fig : illustration }. Then, building on these matrices, we apply our theory of thermodynamic circuits \cite{Raux2024} to obtain the non equilibrium conductance matrix of the serial association of the two CRNs. To conclude this illustration, we verify that the derivation of the non equilibrium conductance matrix from the full network gives the same result. 

This section is organized as follows: we start by describing the modules by providing their stoichiometric matrices and their sub-matrices for internal/external species. Then, we determine the currents and their conjugated forces at all levels of description by exploiting cycles and conservation laws. The modules having pseudo first order dynamics, the stationary concentrations can be computed exactly, either by using the Kirchoff theorem as in appendix B of Ref.~\cite{Avanzini2023} or through purely linear algebraic steps as in Ref.~\cite{raux:tel-04964379}. Using those, the resistance matrix at the reaction level follows. We propagate it to get the conductance matrices at all levels of description. Finally, using the law of resistance matrix addition derived in  \cite{Raux2024}, we determine the conductance matrix for the serial association of modules 1 and 2. We compare our result with the direct computation of the conductance matrix at fundamental level.

\subsection{Stoichiometry and reaction currents}
The stoichiometric matrices describing the reactions for modules 1 and 2 read
\begin{align}
\bm \nabla^{(1)}&=
\begin{bmatrix}
\bm \nabla^{(1)}_\textsc{x}\\
\bm \nabla^{(1)}_\textsc{y}
\end{bmatrix}=
\left[\begin{array}{ccc}
-1 & 1 & 0\\
1 & -1 & -1\\
0 & 0 & 1\\
\hline
-1 & 0 & -1\\
0 & 1 & 0
\end{array}\right],\\
\bm \nabla^{(2)}&=
\begin{bmatrix}
\bm \nabla^{(2)}_\textsc{x}\\
\bm \nabla^{(2)}_\textsc{y}
\end{bmatrix}=
\left[\begin{array}{ccccc}
-1 & 0 & 1 & 0 & 0\\
1 & -1 & 0 & -1 & 0\\
0 & 1 & -1 &0 & 1\\
0 & 0 & 0 & 1 & -1\\
\hline
0 & 0 & 0 & -1 & 0\\
0 & 0 & 0 & 0 & 1\\
-1 & 0 & 0 & 0 & 0\\
0 & 0 & 1 & 0 & 0
\end{array}\right].
\end{align}
The columns (reaction number $\rho$) are ordered respectively as 
\begin{equation}
	\mathscr{R}^{(1)}=\lbrace 1, 2,3 \rbrace, \qquad
	\mathscr{R}^{(2)}=\lbrace 4, 5, 6, 7, 8 \rbrace.
\end{equation}
The rows (species labels $Z_\alpha$) are respectively ordered as
\begin{eqnarray}
	\mathscr{S}^{(1)}&=&\lbrace E_a, E_aS,  E_aS_2, S, N_a \rbrace, \\
	\mathscr{S}^{(2)}&=& \lbrace E_b, E_bF, E_bW, E_b^*, N_a, N_b, F, W\rbrace.
\end{eqnarray} 
The internal and external species are 
\begin{align}
	\mathscr{S}^{(1)}_\textsc{x}=&\lbrace E_a, E_aS,  E_aS_2 \rbrace, &
	\mathscr{S}^{(2)}_\textsc{x}=&  \lbrace E_b, E_bF, E_bW, E_b^*\rbrace, \nonumber\\
	\mathscr{S}^{(1)}_\textsc{y}=&\lbrace S, N_a \rbrace, & 
	\mathscr{S}^{(2)}_\textsc{y}=& \lbrace N_a, N_b, F, W\rbrace. \label{chemostated-species}
\end{align} The horizontal line in the stoichiometric matrices separates them into two submatrices: the upper block $\bm \nabla^{(m)}_\textsc{x}$ and the lower block $\bm \nabla^{(m)}_\textsc{y}$ for $m=1,2$. In the next two sections, we express the stationary currents at all levels of description in terms of the reaction currents 
\begin{align}
\bm j^{(1)}&=
\begin{pmatrix}
j_1\\
j_2\\
j_3
\end{pmatrix}=
\begin{pmatrix}
k_{1}^+ S \left[E_a\right] - k_{1}^- \left[E_aS\right]\\
k_{2}^+ \left[E_a S\right] - k_{2}^- N_a \left[E_a\right]\\
k_{3}^+ S\left[E_aS \right] - k_{3}^- \left[E_aS_2 \right]
\end{pmatrix}, \label{currents 1} \\
\bm j^{(2)}&=
\begin{pmatrix}
j_4\\
j_5\\
j_6\\
j_7\\
j_8
\end{pmatrix}=
\begin{pmatrix}
k_{4}^+ F \left[ E_b\right] - k_{4}^- \left[E_bF\right]\\
k_{5}^+ \left[E_bF\right] - k_{5}^- \left[E_bW\right]\\
k_{6}^+  \left[E_bW\right] - k_{6}^- W  \left[E_b\right]\\
k_{7}^+ N_a  \left[E_bF\right] - k_{7}^-  \left[E_b^*\right]\\
k_{8}^+ \left[E_b^*\right] - k_{8}^- N_b \left[E_bW\right]
\end{pmatrix}, \label{currents 2}
\end{align}
where the concentrations of external species are denoted without square brackets to emphasize that they are parameters for the effective rate constants.

\subsection{From reaction to cycle currents}
As aforementioned, the reaction currents are linearly dependent: A basis of linearly independent cycle currents can thus be chosen. Looking for a basis of the kernel of $\bm \nabla^{(m)}_\textsc{x}$ for $m=1,2$, we choose the following cycle matrices
\begin{equation}
\bm {\mathcal{C}}^{(1)}=
\begin{pmatrix}
1 \\
1\\
0
\end{pmatrix}, \qquad
\bm {\mathcal{C}}^{(2)}=
\begin{bmatrix}
0 & 1\\
-1 & 1\\
0 & 1\\
1 & 0\\
1 & 0
\end{bmatrix}.
\end{equation}
Applying Eq.~\eqref{eq : redundant reaction currents} to both modules yields the following relation between their reaction currents:
\begin{align}
j_1&=j_2,\quad j_3=0,\label{eq : reaction currents dependances A}\\
j_4&=j_6, \quad j_5=j_6-j_8,\quad j_7=j_8.\label{eq : reaction currents dependances B}
\end{align}
We remark that equation $j_3=0$ is due to our assumption of  stationary state: a non zero current would lead to an accumulation of $E_a S_2$ in the system.
We notice also that these relations between reaction currents are an instance of Kirchhoff's current law applied to each species of the CRN. Then, since the cycle matrices are pseudo-invertible ($\bm {\mathcal{C}}^{(1)}$ is a vector and $\bm {\mathcal{C}}^{(2)}$ has linearly independent columns), we can invert  Eq.~\eqref{eq : def cycle currents} for both modules and obtain the following cycle currents 
\begin{align}
J^{(1)}&=\bm {\mathcal{C}}^{(1)+}\bm j^{(1)}=j_1,\label{eq : cycle currents example 1}\\
\bm J^{(2)}&=\bm {\mathcal{C}}^{(2)+}\bm j^{(2)}=
\begin{pmatrix}
j_8\\
j_6
\end{pmatrix}, \label{eq : cycle currents example 2}
\end{align}
compatible with $\bm j^{(m)} = \bm {\mathcal{C}}^{(m)} \bm J^{(m)}$ by using the constraints of Eqs.~(\ref{eq : reaction currents dependances A}--\ref{eq : reaction currents dependances B}).

From the cycle matrices and the stochiometric submatrix for external species, we find the matrices of physical exchanges
\begin{equation}
	\bm \phi^{(1)} = 
	\begin{pmatrix}
		-1 \\ 1
	\end{pmatrix}, \qquad
	\bm \phi^{(2)} = 
	\begin{bmatrix}
		-1& 0 \\
		 1& 0 \\
		 0&-1 \\
		 0& 1  
	\end{bmatrix},	\label{physical-exchanges}
\end{equation}
whose lines are associated to external species as ordered in Eq.~\eqref{chemostated-species}.
\subsection{From physical to fundamental currents}
The left null eigenvectors of $\bm \nabla^{(m)}$ for $m=1,2$ constitute the rows of matrices
\begin{eqnarray}
\bm L^{(1)} &=&
\left[\begin{array}{ccc|cc}
1 & 1 & 1 & 0 & 0\\ \hline
-2 & -1 & 0 & 1 & 1
\end{array}\right] 
= 
\left[ \begin{array}{cc}
\bm \ell^{(1)u}_\textsc{x} & \bm 0 \\
\bm \ell^{(1)b}_\textsc{x} & \bm \ell^{(1)}
\end{array} \right], \\
\bm L^{(2)}&=&\left[
\begin{array}{cccc|cccc}
1 & 1 & 1 & 1 & 0 & 0 & 0 & 0 \\
\hline
-1 & -1 &-1 & 0 & 1 & 1 & 0 & 0\\
-1 & 0 & 0 & 0 & 0 & 0 & 1 & 1
\end{array}\right] = \left[\begin{array}{cc}
\bm \ell^{(2)u}_\textsc{x} & \bm 0 \\
\multirow{ 2}{*}{$\bm \ell_\textsc{x}^{(2)b}$}  & \multirow{ 2}{*}{$\bm \ell^{(2)}$}  \\ &
\end{array} \right]. \nonumber\\
\end{eqnarray}
We separate the unbroken and broken conservation laws (respectively internal and external species) by an horizontal (respectively vertical) line. 
We identify the matrices $\bm \ell^{(m)}$ of conservation laws for the chemical currents received from the chemostats
\begin{equation}
\bm \ell^{(1)}=
\left(\begin{array}{cc}
1 & 1
\end{array}\right), \qquad
\bm \ell^{(2)}=\left[
\begin{array}{cccc}
1 & 1 & 0 & 0\\
0 & 0 & 1 & 1
\end{array}\right].\label{eq : conservation laws 2}
\end{equation}
Those matrices are in the cokernel of the matrices of physical exchanges given in Eq.~\eqref{physical-exchanges}.
The physical currents follow from Eq.~\eqref{eq : definition des courants de chemostats} and read for both modules
\begin{align}
\bm i^{(1)}&=-\bm \nabla_\textsc{y}^{(1)}\bm j^{(1)}=
\begin{pmatrix}
j_1\\
-j_1
\end{pmatrix},\\
\bm i^{(2)}&= - \bm \nabla_\textsc{y}^{(2)}\bm j^{(2)}=
\begin{pmatrix}
j_8\\
-j_8\\
j_6\\
-j_6
\end{pmatrix}.
\end{align}
Their components follow the order of external species in $\mathscr{S}_\textsc{y}^{(m)}$ of Eq.~\eqref{chemostated-species}. They are chosen positive when matter flows from the chemostat to the module as one can check on Fig.~\ref{fig : effective modules}.
Finally, we can select linearly independent currents called fundamental currents by choosing a vector basis for the kernels of the conservation law matrices in Eq.~(\ref{eq : conservation laws 2}). This leads to selection matrices %
\begin{equation}
\bm S^{(1)}=
\begin{pmatrix}
1\\
-1
\end{pmatrix}, \qquad
\bm S^{(2)}=
\begin{bmatrix}
1 & 0\\
-1 & 0\\
0 & 1\\
0 & -1
\end{bmatrix}.\label{eq : selection matrix}
\end{equation}
On can check Eq.~\eqref{eq : fundamental chemostats currents} with the above selection matrices for the following fundamental currents
\begin{equation}
I^{(1)}=j_1, \qquad
\bm I^{(2)}=
\begin{pmatrix}
j_8\\
j_6
\end{pmatrix}.
\end{equation}
In the end, module 1 is described by an effective reaction converting $S$ into $N_a$ and one reaction current $j_1$. Similarly, module 2 is described by two coupled effective reactions and two reaction currents $j_6$ and $j_8$. Fig.~\ref{fig : effective modules} summarizes this effective description for modules 1 and 2 in serial association \cite{Avanzini2023}. 
\subsection{Thermodynamic forces }
We now turn to the derivation of the conjugated thermodynamics forces in terms of the reaction affinities defined in Eq.~\eqref{eq : reaction affinity} and which read for modules 1 and 2
\begin{align}
\bm f^{(1)}&=
\begin{pmatrix}
f_1\\
f_2\\
f_3
\end{pmatrix}=
\begin{pmatrix}
\mu_{E_a}+\mu_S-\mu_{E_aS} \\
\mu_{E_aS}-\mu_{N_a}-\mu_{E_a}\\
\mu_{E_aS}+\mu_S-\mu_{E_aS_2}
\end{pmatrix},\label{eq : reaction affinities exemple 1}\\
\bm f^{(2)}&=
\begin{pmatrix}
f_4\\
f_5\\
f_6\\
f_7\\
f_8
\end{pmatrix}=
\begin{pmatrix}
\mu_{E_b}+\mu_F-\mu_{E_bF}\\
\mu_{E_bF}-\mu_{E_bW}\\
\mu_{E_bW}-\mu_W-\mu_{E_b}\\
\mu_{E_bF} + \mu_{N_a} - \mu_{E_b^*}\\
\mu_{E_b^*}-\mu_{N_b}-\mu_{E_bW} 
\end{pmatrix}.\label{eq : reaction affinities exemple 2}
\end{align}
Using these reaction affinities for modules 1 and 2 in the definition of cycle forces Eq.~\eqref{eq : definition of the cycle affinity}, we obtain
\begin{align}
F^{(1)}&={\bm {\mathcal{C}}^{(1)}}^T \bm f^{(1)}=f_1+f_2=\mu_S-\mu_{N_a},\label{eq : cycle affinity 1}\\
\bm F^{(2)}&={\bm {\mathcal{C}}^{(2)}}^T \bm f^{(2)}=\begin{pmatrix}
f_7+f_8-f_5\\
f_4+f_5+f_6
\end{pmatrix}=\begin{pmatrix}
\mu_{N_a}-\mu_{N_b}\\
\mu_{F}-\mu_{W}
\end{pmatrix}.
\end{align}
Those cycle forces are conjugated to the currents of Eqs.~(\ref{eq : cycle currents example 1}--\ref{eq : cycle currents example 2}). The physical forces are the chemical potential associated to the external species 
\begin{align}
\bm a^{(1)T}&=\begin{pmatrix}
\mu_S & \mu_{N_a}
\end{pmatrix},\label{eq : chemostat affinity 1}\\
\bm a^{(2)T}&=
\begin{pmatrix}
\mu_{N_a} & \mu_{N_b} & \mu_F & \mu_{W}
\end{pmatrix}.\label{eq : chemostat affinity 2}
\end{align}
Finally, the fundamental forces are obtained by applying the definition of Eq.~\eqref{eq : fundamental affinities} with the selection matrices of Eq.~(\ref{eq : selection matrix}) and the physical forces of Eqs.~(\ref{eq : chemostat affinity 1}--\ref{eq : chemostat affinity 2})
\begin{align}
A^{(1)T}&=\bm a^{(1)T}\bm S^{(1)}=\mu_{S}-\mu_{N_a},\\
\bm A^{(2)T}&=\bm a^{(2)T} \bm S^{(2)}=\begin{pmatrix}
\mu_{N_a}-\mu_{N_b} & \mu_F-\mu_W
\end{pmatrix}.
\end{align}
Note that with this choice of selection matrices, the fundamental forces for both modules are equal to the cycle forces. 
\subsection{Conductance matrices}
We compute now the resistance and conductance matrices at all levels for module 1 and 2. First, at the reactions level, the resistance matrices are denoted
\begin{align}
\bm r^{(1)}&=\begin{bmatrix}
r_1 & 0 & 0\\
0    & r_2 & 0\\
0 &  0  & r_3
\end{bmatrix},\\
\bm r^{(2)}&=
\begin{bmatrix}
r_4 & 0 & 0 & 0 & 0\\
0 & r_5 & 0 & 0 & 0\\
0 & 0 & r_6 & 0 & 0\\
0 & 0 & 0 & r_7 & 0\\
0 & 0 & 0 & 0 & r_8
\end{bmatrix}.
\end{align}
with the $r_{\rho}$ obtained from Eqs.~(\ref{currents 1}--\ref{currents 2}) and (\ref{eq : reaction affinities exemple 1}--\ref{eq : reaction affinities exemple 2}).
Using Eq.~\eqref{eq : cycle resistance matrix}, the cycle resistance matrices for each module read
\begin{align}
 R^{(1)}&=
r_1+r_2,\\
\bm R^{(2)}&=\begin{bmatrix}
r_5+r_7+r_8 & -r_5\\
-r_5 & r_4+r_5+r_6 
\end{bmatrix}.
\end{align}
As expected, since there is a single cycle current for module 1, the cycle resistance is scalar. 
Module 2 has two cycles, its resistance matrix is thus a $2\times 2$ matrix. Its diagonal elements displays the resistance addition of the reaction involved in each cycles. Its off-diagonal elements characterize the coupling between the transport of chemical species along the cycles pathways. Looking at the CRN of module 2 on Fig.~\ref{fig : illustration }, one expects that reaction for $\rho=5$ couples the two reaction cycles. 
Using equation Eq.~\eqref{eq : chemostat conductance matrix}, the conductance matrices at physical level read
\begin{align}
\bm g^{(1)}&=\frac{1}{r_1+r_2}\;\bm{\mathcal{U}} \quad \text{ with } \quad \bm{\mathcal{U}}=\begin{bmatrix}
1 & -1\\
-1 & 1
\end{bmatrix}, \\[0.5em]
\bm g^{(2)}&=\frac{1}{\det \bm R^{(2)}}\begin{bmatrix}
 (r_4+r_5+r_6)\; \bm{\mathcal{U}} & r_5\; \bm{\mathcal{U}}\\
 r_5 \; \bm{\mathcal{U}}&  \left(r_5+r_7+r_8\right)\; \bm{\mathcal{U}}
\end{bmatrix}.
\end{align}
Finally, the conductance matrices at fundamental level are obtained from Eq.~\eqref{eq : fundamental conductance matrix}: 
\begin{align}
 G^{(1)}&=\frac{1}{r_1+r_2},\\[0.5em]
\bm G^{(2)}&=\frac{1}{\det \bm R^{(2)}}
\begin{bmatrix}
r_4+r_5+r_6& r_5\\
r_5 & r_5+r_7+r_8
\end{bmatrix}.
\end{align} 
In other words the cycle and fundamental resistance matrices are equal for this CRN. 
\subsection{Conductance matrix of the full network}
We refer to module 3 as the serial association of modules 1 and 2 via the chemical species $N_a$. Therefore, $N_a$, which was an external species for modules 1 and 2 when they were studied separately, is now an internal species from the point of view of module 3. In this section, we present two alternative derivations of the non-equilibrium conductance matrix of module 3. First, we explain how our law of resistance addition for serially connected thermodynamic devices applies to this example. Then, we apply the procedure developed in this paper to the larger network of module 3. We obtain the same result from the two derivations.
\subsubsection{Law of resistance matrix addition}
In the previous sections, module 1 and 2 are studied separately, i.e., as if reactions occur into separate reactors. We now turn to the serial association of modules 1 and 2 to give module 3. Upon connection, all reactions occur in the same reactor with conservation of the chemical current corresponding to exchanges of $N_a$ between the two modules. Moreover, the concentration $\left[N_a\right]$ at the ``interface'' between the modules reaches a unique nonequilibrium stationary state. In other words, the serial connection implies $j_1=j_8$ and chemical potential continuity at the module's ``interface'' for $N_a$. 

On this basis, we can apply our general method to compute the non equilibrium conductance matrix of module 3, see Ref.~\cite{Raux2024}. As a start, we determine the conservation laws of module 3 and provide the currents at the interface between the two modules as a function of the chemostat currents of module 3. 
The left/right splitting of physical currents reads
\begin{align}
\bm i^{(1)}&=\left(\begin{array}{c}
i_l^{(1)}\\ 
i_r^{(1)}
\end{array}\right)=
\left(\begin{array}{c}
j_1 \\
-j_1
\end{array}\right),\label{eq : left right chemostat currents 1}\\
\bm i^{(2)}&=\left(
\begin{array}{c}
 i_l^{(2)}\\
\bm i_r^{(2)}
\end{array}\right)=\left(
\begin{array}{c}
j_8\\
-j_8\\
j_6\\
-j_6
\end{array}\right),\label{eq : left right chemostats currents 2}
\end{align}
where $\bm i_r^{(2)}$ is the vector made with the last three components of $\bm i^{(2)}$. By construction, the physical currents from the chemostats of module 3 are
\begin{equation}
\bm i^{(3)} = \begin{pmatrix}
 i_l^{(1)}\\ 
\bm i_r^{(2)}
\end{pmatrix}=
\begin{pmatrix}
j_1\\
-j_8\\
j_6\\
-j_6
\end{pmatrix},
\end{equation}
since the current leaving module 1 from the right and entering module 2 on the left are eliminated from our description after serial connection of the two modules.
The conservation laws of Eq.~(\ref{eq : conservation laws 2}) can be reformulated to separate interface and chemostat currents for module 3. In the equation below, we place on the left hand side the interface currents and on the right hand side the chemostat currents:
\begin{equation}
\bm L_i i_r^{(1)}=\bm L_e
\bm i^{(3)},
\label{eq : serial core equation}
\end{equation}
with 
\begin{equation}
\bm L_i = \begin{bmatrix}
-1\\
1\\
0
\end{bmatrix}, \quad \bm L_e =
\begin{bmatrix}
1 & 0 & 0 & 0\\
0 & 1 & 0 & 0\\
0 & 0 & 1 & 1
\end{bmatrix},
\end{equation}
the matrices resulting from this transformation of conservation laws [see Eq.~(40) in Ref.~\cite{Raux2024}]. Eq.~\eqref{eq : serial core equation} allows us to obtain the conservation laws for module 3: we denote
\begin{equation}
\bm v = \begin{bmatrix}
1 & 1 & 0\\
0 & 0 & 1
\end{bmatrix}
\end{equation}
the matrix whose lines are basis vectors of the cokernel of $\bm L_i$ [see Eq.~(45) in Ref.~\cite{Raux2024}]. Left multiplying Eq.~\eqref{eq : serial core equation} by $\bm v$ produces the conservation laws for physical currents of module 3 
\begin{equation}
\bm \ell^{(3)}\bm i^{(3)}=\bm 0,
\end{equation}
where 
\begin{equation}
\bm \ell^{(3)}=\bm v \bm L_e=\begin{bmatrix}
1 & 1 & 0 & 0\\
0 & 0 & 1 & 1
\end{bmatrix}.
\label{eq : conservation laws 3}
\end{equation}
It turns out that the conservation law matrices for module 2 and 3 are equal. This is expected since a single pin connection doesn't decrease the number of external species. In practice, module 1 puts one pin of module 2 in mixed boundary conditions. A choice of selection matrix associated to the conservation law matrix of Eq.~\eqref{eq : conservation laws 3} is thus
\begin{equation}
\bm S^{(3)}=\bm S^{(2)}.
\label{eq : selection matrix 3}
\end{equation}
In the same line, the equivalent description of module 3 is very similar to the one of module 2 as shown on Fig.~\ref{fig : effective modules}.
Going back to Eq.~\eqref{eq : serial core equation}, we can solve for $ i_r^{(1)}$ to express it in terms of $\bm i^{(3)}$ as $i_r^{(1)}=\bm \pi \bm i^{(3)} = - i_{l}^{(2)}$ with
\begin{equation}
\bm \pi =\bm L_i^+\bm L_e =\frac{1}{2}\begin{bmatrix}
-1 & 1 & 0 & 0 
\end{bmatrix}.
\end{equation}
Inserting this last relation in Eqs.~(\ref{eq : left right chemostat currents 1}-- \ref{eq : left right chemostats currents 2}) yields the following relation between the physical currents of modules 1 and 2 in terms of those of module 3
\begin{align}
\bm i^{(m)}=\bm \pi^{(m,3)} \bm i^{(3)}
\label{eq : im function of i3}
\end{align}
for $m=1,2$ and where 
\begin{align}
\bm \pi^{(1,3)}& = \begin{bmatrix}
1 & 0 & 0 & 0\\
-\frac{1}{2} & \frac{1}{2} & 0 & 0
\end{bmatrix},\\
\bm \pi^{(2,3)}&=
\begin{bmatrix}
\frac{1}{2} & -\frac{1}{2} & 0 & 0\\
0 & 1 & 0 & 0\\
0 & 0 & 1 & 0\\
0 & 0 & 0 & 1
\end{bmatrix}.
\end{align}
Finally, the relation between the fundamental currents of module 1 and 2 and those of module 3 is obtained by using Eq.~\eqref{eq : fundamental chemostats currents} with the selection matrix of Eq.~\eqref{eq : selection matrix 3} in Eq.~\eqref{eq : im function of i3} 
\begin{equation}
\bm i^{(m)}=\bm \pi^{(m,3)} \bm S^{(3)} \bm I^{(3)}.
\end{equation}
Then, since the selection matrices for module 1 and 2 are pseudo-invertible, we finally obtain
\begin{equation}
\bm I^{(m)}=\bm \Pi^{(m,3)}\bm I^{(3)}
\end{equation}
with $\bm \Pi^{(m,3)} \equiv \bm S^{(m)+}\bm \pi^{(m,3)}\bm S^{(3)} $ giving
\begin{equation}
\bm \Pi^{(1,3)}=\begin{bmatrix}
1 & 0
\end{bmatrix} \; \text{ and } \; 
\bm \Pi^{(2,3)}= \begin{bmatrix} 1 & 0 \\ 0 & 1 \end{bmatrix}.
\end{equation}
Using the additive structure for resistance matrices [see Eq.~(34) in Ref.~\cite{Raux2024}]  
\begin{equation}
    {\bm G^{(3)}}^{-1}=\sum_{m=1}^2 {\bm \Pi^{(m,3)}}^T{\bm G^{(m)}}^{-1}{\bm \Pi^{(m,3)}},
\end{equation}
we get the fundamental resistance matrix of module 3
\begin{equation}
{\bm G^{(3)}}^{-1}=\begin{bmatrix}
r_1+r_2+r_5+r_7+r_8 & -r_5\\
-r_5 & r_4+r_5+r_6
\end{bmatrix}
\label{eq : fundamental conductance matrix CRN 3 1}
\end{equation}
describing the force-current characteristics. Interestingly the coupling between the fundamental currents for module 3, i.e., the off diagonal elements in ${\bm G^{(3)}}^{-1}$, emerges solely from the chemical reaction coupling the two cycles of module 2. The intensity  of the first diagonal element is the sum of the resistance of the two cycles involved in the serial association. The second diagonal coefficient is equal to the cycle resistance of the cycle that is not involved in the serial association. 
\subsubsection{Direct derivation}
In this section, we provide an alternative derivation of the conductance matrix of Eq.~\eqref{eq : fundamental conductance matrix CRN 3 1} for module 3. We compute this matrix directly from the stoichiometric matrix of the entire CRN. This latter reads
\begin{equation}
\bm \nabla^{(3)}=
\begin{bmatrix}
\bm \nabla_\textsc{x}^{(3)}\\
\bm \nabla_\textsc{y}^{(3)}
\end{bmatrix}=
\left[
\begin{array}{cccccccc}
-1 & 1 & 0 & 0 & 0 & 0 & 0 & 0\\
1 & -1 & -1 & 0 & 0 & 0 & 0 & 0\\
0 & 0 & 1 & 0 & 0 & 0 & 0 & 0\\
0 & 1 & 0 & 0 & 0 & 0 & -1 & 0\\
0 & 0 & 0 & -1 & 0 & 1 & 0 & 0\\
0 & 0 & 0 & 1 & -1 & 0 & -1 & 0\\
0 & 0 & 0 & 0 & 1 & -1 & 0 & 1\\
0 & 0 & 0 & 0 & 0 & 0 & 1 & -1\\
\hline
-1 & 0 & -1 & 0 & 0 & 0 & 0 & 0\\
0 & 0 & 0 & 0 & 0 & 0 & 0 & 1\\
0 & 0 & 0 & -1 & 0 & 0 & 0 & 0\\
0 & 0 & 0 & 0 & 0 & 1 & 0 & 0
\end{array}\right].
\end{equation}
In this matrix, the columns are numbered by the reaction index $\rho=1,\dots, 8$. The rows correspond to the chemical species ordered as in 
\begin{multline}
    \mathscr{S}^{(3)}= \lbrace E_a, E_aS,  E_aS_2, N_a, \\E_b, E_bF, E_bW, E_b^*, S, N_b, F, W\rbrace.
\end{multline}
The upper left $3\times3$ submatrix is $\bm \nabla_\textsc{x}^{(1)}$. The lower right $4\times 5$ submatrix above the horizontal line is $\bm \nabla_\textsc{x}^{(2)}$. The lower right $3 \times 5$ submatrix is $\bm \nabla_\textsc{y}^{(2)}$. The fourth row that corresponds to $N_a$ appears now in the internal species (upper part) of the stoichiometric matrix. The ninth row that correspond to $S$ remains in the external species (lower part).
The whole networks admits the two following cycles 
\begin{equation}
\bm {\mathcal{C}}^{(3)}=
\left[
\begin{array}{cc}
 1 & 0 \\
 1 & 0 \\
 0 & 0 \\
 0 & 1 \\
 -1 & 1 \\
 0 & 1 \\
 1 & 0 \\
 1 & 0 \\
\end{array}
\right].
\end{equation}
The reaction resistance matrix is the diagonal matrix:
\begin{equation}
\bm r^{(3)}=\mathrm{diag} (r_1, r_2, r_3, r_4, r_5, r_6, r_7, r_8) .
\end{equation}
The cycle resistance matrix thus reads 
\begin{equation}
\bm R^{(3)}=\left[
\begin{array}{cc}
 r_1+r_2+r_5+r_7+r_8 & -r_5 \\
 -r_5 & r_4+r_5+r_6 \\
\end{array}
\right].
\end{equation}
To obtain the conservation laws, we look for the cokernel of the stoichiometric matrix which takes the form 
\begin{equation}
\bm L^{(3)}=
\left[
\begin{array}{cccccccc|cccc}
 0 & 0 & 0 & 0 & 1 & 1 & 1 & 1 & 0 & 0 & 0 & 0 \\
 1 & 1 & 1 & 0 & 0 & 0 & 0 & 0 & 0 & 0 & 0 & 0 \\ \hline
 -2 & -1 & 0 & 1 & -1 & -1 & -1 & 0 & 1 & 1 & 0 & 0 \\
 0 & 0 & 0 & 0 & -1 & 0 & 0 & 0 & 0 & 0 & 1 & 1 \\
  \end{array}
\right].
\end{equation} 
The two last line correspond to the broken conservation laws since they have non zero coefficients for the external species. The conservation laws for chemostat currents are recovered as the lower right block in $\bm L^{(3)}$. We emphasize that we chose for $\bm L^{(3)}$ an appropriate row ordering of the left null eigenvectors so as to get a matrix of conservation laws in the form of Eq.~\eqref{Cannonical-Cons-Laws}. 
Given that matrix of current conservation law is identical to the one found in the serial approach, the same choice of selection matrix can be made leading to the fundamental resistance matrix given in Eq.~\eqref{eq : fundamental conductance matrix CRN 3 1}. Therefore, we conclude that, for the simple CRNs studied in this section, the serial connection of chemical modules leads to the same results as the direct approach on the full network.

\section{Conclusion}
In this work, we have defined (for the four levels of description: reaction, cycle, physical and fundamental) the nonequilibrium conductance matrix of a CRN in its (unique) stationary nonequilibrium state. This definition involves the reaction cycles for internal species, the matter exchanged with the chemostats for each cycle and the conservation laws of the CRN leading to a choice of selection matrix associated to a given basis of fundamental currents and forces. Once these concepts have been identified in the framework of CRNs, the definition for the conductance matrix is analogous to the one for Markov jump processes \cite{Vroylandt2018, Raux2024}. 
This approach leads to a synthetic description of a CRN in terms of its nonequilibrium conductance matrix and conservation laws for physical currents, in the spirit of Ref.~\cite{Raux2024}, as compared to the approach based on emergent cycles of Ref.~\cite{Polettini2014, Avanzini2023}. 

The nonequilibrium conductance matrix is uniquely defined when  assuming that the CRN reaches a unique nonequilibrium stationary state. This is guaranteed for pseudo-first order kinetics which are linear, although we remark that the notion of linearity depends on the decomposition of the CRN into chemical modules. It may happen that non-linearity reappears upon connection of linear modules, as it is the case for our illustrative example. In this case, the problem of non-uniticity may reappear when solving for the concentration of the species at the module's interface. This problem of multiple solutions deserves further investigations, for instance regarding stability criteria. Borrowing from the theory of electronic circuits will certainly be useful in this direction, all the more so for emergent phenomena commonly appearing in non-linear systems. This may help to advance on the description of  nonequilibrium phase transitions given the crucial lack of nonequilibrium thermodynamic potentials and associated variational principles.

\appendix

\section{Relative dimension of cycle and fundamental forces}
\label{fund or cycle forces}

The main text examples has cycle and fundamental vectors that share the same dimension. We show in this appendix that this is not the case when $\mathrm{ker} \,( \bm \nabla) $ is not null. The proof relies on several uses of the rank nullity theorem for matrices $ \bm \nabla_\textsc{x}$, $ \bm \nabla$ and their transposes. This theorem states that the number of columns of a matrix is equal to the rank of the matrix plus the dimension of its kernel, i.e.,
\begin{eqnarray}
	|\mathscr{R}|&=& \mathrm{rk}\, \bm \nabla_\textsc{x} + |\mathscr{C}|,\\
|\mathscr{S}_\textsc{x}|  &=& \mathrm{rk}\, \bm \nabla_\textsc{x}^{T}+ |\mathscr{L}^{u}|,\\
	|\mathscr{R}|&=& \mathrm{rk}\, \bm \nabla + \mathrm{dim}(\mathrm{ker}\, \bm \nabla),\\
	|\mathscr{S}|  &=& \mathrm{rk}\, \bm \nabla^{T} + |\mathscr{L}_\mathrm{cl}|.
\end{eqnarray} 
Since a matrix and its transpose have identical rank, we obtain
\begin{eqnarray}
	|\mathscr{C}| &=&-|\mathscr{S}_\textsc{x}| +|\mathscr{L}^{u}|+|\mathscr{R}|, \\
	&=& |\mathscr{S}_\textsc{y}| -|\mathscr{L}^{b}|+|\mathscr{L}_\mathrm{cl}|-|\mathscr{S}| + |\mathscr{R}|, \\
	&=&|\mathscr{S}_\textsc{y}| -|\mathscr{L}^{b}| + \mathrm{dim}(\mathrm{ker}\, \bm \nabla).
\end{eqnarray}
Hence, given that $|\mathscr{C}|$ is the number of cycle forces and $|\mathscr{S}_\textsc{y}| -|\mathscr{L}^{b}|$ the number of fundamental forces, the two are different for non-zero dimensions of the stochiometric matrix's kernel.

\bibliography{biblio_chimie}

\begin{thebibliography}{20}%
\makeatletter
\providecommand \@ifxundefined [1]{%
 \@ifx{#1\undefined}
}%
\providecommand \@ifnum [1]{%
 \ifnum #1\expandafter \@firstoftwo
 \else \expandafter \@secondoftwo
 \fi
}%
\providecommand \@ifx [1]{%
 \ifx #1\expandafter \@firstoftwo
 \else \expandafter \@secondoftwo
 \fi
}%
\providecommand \natexlab [1]{#1}%
\providecommand \enquote  [1]{``#1''}%
\providecommand \bibnamefont  [1]{#1}%
\providecommand \bibfnamefont [1]{#1}%
\providecommand \citenamefont [1]{#1}%
\providecommand \href@noop [0]{\@secondoftwo}%
\providecommand \href [0]{\begingroup \@sanitize@url \@href}%
\providecommand \@href[1]{\@@startlink{#1}\@@href}%
\providecommand \@@href[1]{\endgroup#1\@@endlink}%
\providecommand \@sanitize@url [0]{\catcode `\\12\catcode `\$12\catcode
  `\&12\catcode `\#12\catcode `\^12\catcode `\_12\catcode `\%12\relax}%
\providecommand \@@startlink[1]{}%
\providecommand \@@endlink[0]{}%
\providecommand \url  [0]{\begingroup\@sanitize@url \@url }%
\providecommand \@url [1]{\endgroup\@href {#1}{\urlprefix }}%
\providecommand \urlprefix  [0]{URL }%
\providecommand \Eprint [0]{\href }%
\providecommand \doibase [0]{http://dx.doi.org/}%
\providecommand \selectlanguage [0]{\@gobble}%
\providecommand \bibinfo  [0]{\@secondoftwo}%
\providecommand \bibfield  [0]{\@secondoftwo}%
\providecommand \translation [1]{[#1]}%
\providecommand \BibitemOpen [0]{}%
\providecommand \bibitemStop [0]{}%
\providecommand \bibitemNoStop [0]{.\EOS\space}%
\providecommand \EOS [0]{\spacefactor3000\relax}%
\providecommand \BibitemShut  [1]{\csname bibitem#1\endcsname}%
\let\auto@bib@innerbib\@empty
\bibitem [{\citenamefont {Schnakenberg}(1976)}]{Schnakenberg1976}%
  \BibitemOpen
  \bibfield  {author} {\bibinfo {author} {\bibfnamefont {J.}~\bibnamefont
  {Schnakenberg}},\ }\href {\doibase 10.1103/RevModPhys.48.571} {\bibfield
  {journal} {\bibinfo  {journal} {Rev. Mod. Phys.}\ }\textbf {\bibinfo {volume}
  {48}},\ \bibinfo {pages} {571} (\bibinfo {year} {1976})}\BibitemShut
  {NoStop}%
\bibitem [{\citenamefont {Hill}(1989)}]{Hill1989}%
  \BibitemOpen
  \bibfield  {author} {\bibinfo {author} {\bibfnamefont {T.~L.}\ \bibnamefont
  {Hill}},\ }\href {\doibase 10.1007/978-1-4612-3558-3} {\emph {\bibinfo
  {title} {Free Energy Transduction and Biochemical Cycle Kinetics}}}\
  (\bibinfo  {publisher} {Springer-Verlag New York, Inc.},\ \bibinfo {year}
  {1989})\BibitemShut {NoStop}%
\bibitem [{\citenamefont {Dal~Cengio}\ \emph {et~al.}(2023)\citenamefont
  {Dal~Cengio}, \citenamefont {Lecomte},\ and\ \citenamefont
  {Polettini}}]{DalCengio2023}%
  \BibitemOpen
  \bibfield  {author} {\bibinfo {author} {\bibfnamefont {S.}~\bibnamefont
  {Dal~Cengio}}, \bibinfo {author} {\bibfnamefont {V.}~\bibnamefont {Lecomte}},
  \ and\ \bibinfo {author} {\bibfnamefont {M.}~\bibnamefont {Polettini}},\
  }\href {\doibase 10.1103/PhysRevX.13.021040} {\bibfield  {journal} {\bibinfo
  {journal} {Phys. Rev. X}\ }\textbf {\bibinfo {volume} {13}},\ \bibinfo
  {pages} {021040} (\bibinfo {year} {2023})}\BibitemShut {NoStop}%
\bibitem [{\citenamefont {Ross}(2008)}]{Ross2008}%
  \BibitemOpen
  \bibfield  {author} {\bibinfo {author} {\bibfnamefont {J.}~\bibnamefont
  {Ross}},\ }\href@noop {} {\emph {\bibinfo {title} {Thermodynamics and
  Fluctuations Far from Equilibrium}}},\ edited by\ \bibinfo {editor}
  {\bibfnamefont {S.~R.}\ \bibnamefont {Berry}}\ and\ \bibinfo {editor}
  {\bibfnamefont {K.}~\bibnamefont {Yamanouchi}},\ \bibinfo {series} {Springer
  Series in Chemical Physics Ser.}\ No.\ \bibinfo {number} {v.90}\ (\bibinfo
  {publisher} {Springer Berlin / Heidelberg},\ \bibinfo {address} {Berlin,
  Heidelberg},\ \bibinfo {year} {2008})\ \bibinfo {note} {description based on
  publisher supplied metadata and other sources.}\BibitemShut {Stop}%
\bibitem [{\citenamefont {Van~Kampen}(2007)}]{vanKampen2007}%
  \BibitemOpen
  \bibfield  {author} {\bibinfo {author} {\bibfnamefont {N.}~\bibnamefont
  {Van~Kampen}},\ }\href@noop {} {\emph {\bibinfo {title} {Stochastic Processes
  in Physics and Chemistry}}},\ \bibinfo {edition} {3rd}\ ed.\ (\bibinfo
  {publisher} {North-Holland, Amsterdam},\ \bibinfo {year} {2007})\BibitemShut
  {NoStop}%
\bibitem [{\citenamefont {Nicolis}\ and\ \citenamefont
  {Prigogine}(1977)}]{Nicolis1977}%
  \BibitemOpen
  \bibfield  {author} {\bibinfo {author} {\bibfnamefont {G.}~\bibnamefont
  {Nicolis}}\ and\ \bibinfo {author} {\bibfnamefont {I.}~\bibnamefont
  {Prigogine}},\ }\href@noop {} {\emph {\bibinfo {title} {{Self-Organization in
  Nonequilibrium Systems: From Dissipative Structures to Order Through
  Fluctuations}}}},\ edited by\ \bibinfo {editor} {\bibfnamefont
  {I.}~\bibnamefont {Jon Wiley \&~sons}}\ (\bibinfo  {publisher} {Wiley, New
  York},\ \bibinfo {year} {1977})\BibitemShut {NoStop}%
\bibitem [{\citenamefont {Schmiedl}\ and\ \citenamefont
  {Seifert}(2007)}]{Schmiedl2007}%
  \BibitemOpen
  \bibfield  {author} {\bibinfo {author} {\bibfnamefont {T.}~\bibnamefont
  {Schmiedl}}\ and\ \bibinfo {author} {\bibfnamefont {U.}~\bibnamefont
  {Seifert}},\ }\href {\doibase http://dx.doi.org/10.1063/1.2428297} {\bibfield
   {journal} {\bibinfo  {journal} {J. Chem. Phys.}\ }\textbf {\bibinfo {volume}
  {126}},\ \bibinfo {eid} {044101} (\bibinfo {year} {2007})}\BibitemShut
  {NoStop}%
\bibitem [{\citenamefont {Rao}\ and\ \citenamefont {Esposito}(2016)}]{Rao2016}%
  \BibitemOpen
  \bibfield  {author} {\bibinfo {author} {\bibfnamefont {R.}~\bibnamefont
  {Rao}}\ and\ \bibinfo {author} {\bibfnamefont {M.}~\bibnamefont {Esposito}},\
  }\href {\doibase 10.1103/PhysRevX.6.041064} {\bibfield  {journal} {\bibinfo
  {journal} {Phys. Rev. X}\ }\textbf {\bibinfo {volume} {6}},\ \bibinfo {pages}
  {041064} (\bibinfo {year} {2016})}\BibitemShut {NoStop}%
\bibitem [{\citenamefont {Avanzini}\ \emph {et~al.}(2024)\citenamefont
  {Avanzini}, \citenamefont {Bilancioni}, \citenamefont {Cavina}, \citenamefont
  {Cengio}, \citenamefont {Esposito}, \citenamefont {Falasco}, \citenamefont
  {Forastiere}, \citenamefont {Freitas}, \citenamefont {Garilli}, \citenamefont
  {Harunari}, \citenamefont {Lecomte}, \citenamefont {Lazarescu}, \citenamefont
  {Srinivas}, \citenamefont {Moslonka}, \citenamefont {Neri}, \citenamefont
  {Penocchio}, \citenamefont {Pineros}, \citenamefont {Polettini},
  \citenamefont {Raghu}, \citenamefont {Raux}, \citenamefont {Sekimoto},\ and\
  \citenamefont {Soret}}]{Avanzini2024}%
  \BibitemOpen
  \bibfield  {author} {\bibinfo {author} {\bibfnamefont {F.}~\bibnamefont
  {Avanzini}}, \bibinfo {author} {\bibfnamefont {M.}~\bibnamefont
  {Bilancioni}}, \bibinfo {author} {\bibfnamefont {V.}~\bibnamefont {Cavina}},
  \bibinfo {author} {\bibfnamefont {S.~D.}\ \bibnamefont {Cengio}}, \bibinfo
  {author} {\bibfnamefont {M.}~\bibnamefont {Esposito}}, \bibinfo {author}
  {\bibfnamefont {G.}~\bibnamefont {Falasco}}, \bibinfo {author} {\bibfnamefont
  {D.}~\bibnamefont {Forastiere}}, \bibinfo {author} {\bibfnamefont
  {N.}~\bibnamefont {Freitas}}, \bibinfo {author} {\bibfnamefont
  {A.}~\bibnamefont {Garilli}}, \bibinfo {author} {\bibfnamefont {P.~E.}\
  \bibnamefont {Harunari}}, \bibinfo {author} {\bibfnamefont {V.}~\bibnamefont
  {Lecomte}}, \bibinfo {author} {\bibfnamefont {A.}~\bibnamefont {Lazarescu}},
  \bibinfo {author} {\bibfnamefont {S.~G.~M.}\ \bibnamefont {Srinivas}},
  \bibinfo {author} {\bibfnamefont {C.}~\bibnamefont {Moslonka}}, \bibinfo
  {author} {\bibfnamefont {I.}~\bibnamefont {Neri}}, \bibinfo {author}
  {\bibfnamefont {E.}~\bibnamefont {Penocchio}}, \bibinfo {author}
  {\bibfnamefont {W.~D.}\ \bibnamefont {Pineros}}, \bibinfo {author}
  {\bibfnamefont {M.}~\bibnamefont {Polettini}}, \bibinfo {author}
  {\bibfnamefont {A.}~\bibnamefont {Raghu}}, \bibinfo {author} {\bibfnamefont
  {P.}~\bibnamefont {Raux}}, \bibinfo {author} {\bibfnamefont {K.}~\bibnamefont
  {Sekimoto}}, \ and\ \bibinfo {author} {\bibfnamefont {A.}~\bibnamefont
  {Soret}},\ }\href {\doibase 10.21468/SciPostPhysLectNotes.80} {\bibfield
  {journal} {\bibinfo  {journal} {SciPost Phys. Lect. Notes}\ ,\ \bibinfo
  {pages} {80}} (\bibinfo {year} {2024})}\BibitemShut {NoStop}%
\bibitem [{\citenamefont {Avanzini}\ \emph {et~al.}(2020)\citenamefont
  {Avanzini}, \citenamefont {Falasco},\ and\ \citenamefont
  {Esposito}}]{Avanzini2020}%
  \BibitemOpen
  \bibfield  {author} {\bibinfo {author} {\bibfnamefont {F.}~\bibnamefont
  {Avanzini}}, \bibinfo {author} {\bibfnamefont {G.}~\bibnamefont {Falasco}}, \
  and\ \bibinfo {author} {\bibfnamefont {M.}~\bibnamefont {Esposito}},\ }\href
  {\doibase 10.1088/1367-2630/abafea} {\bibfield  {journal} {\bibinfo
  {journal} {New Journal of Physics}\ }\textbf {\bibinfo {volume} {22}},\
  \bibinfo {pages} {093040} (\bibinfo {year} {2020})}\BibitemShut {NoStop}%
\bibitem [{\citenamefont {Anderson}\ \emph {et~al.}(2015)\citenamefont
  {Anderson}, \citenamefont {Craciun}, \citenamefont {Gopalkrishnan},\ and\
  \citenamefont {Wiuf}}]{Anderson2015}%
  \BibitemOpen
  \bibfield  {author} {\bibinfo {author} {\bibfnamefont {D.~F.}\ \bibnamefont
  {Anderson}}, \bibinfo {author} {\bibfnamefont {G.}~\bibnamefont {Craciun}},
  \bibinfo {author} {\bibfnamefont {M.}~\bibnamefont {Gopalkrishnan}}, \ and\
  \bibinfo {author} {\bibfnamefont {C.}~\bibnamefont {Wiuf}},\ }\href {\doibase
  10.1007/s11538-015-0102-8} {\bibfield  {journal} {\bibinfo  {journal} {Bull.
  Math. Biol.}\ }\textbf {\bibinfo {volume} {77}},\ \bibinfo {pages} {1744}
  (\bibinfo {year} {2015})}\BibitemShut {NoStop}%
\bibitem [{\citenamefont {Avanzini}\ \emph {et~al.}(2023)\citenamefont
  {Avanzini}, \citenamefont {Freitas},\ and\ \citenamefont
  {Esposito}}]{Avanzini2023}%
  \BibitemOpen
  \bibfield  {author} {\bibinfo {author} {\bibfnamefont {F.}~\bibnamefont
  {Avanzini}}, \bibinfo {author} {\bibfnamefont {N.}~\bibnamefont {Freitas}}, \
  and\ \bibinfo {author} {\bibfnamefont {M.}~\bibnamefont {Esposito}},\ }\href
  {\doibase 10.1103/PhysRevX.13.021041} {\bibfield  {journal} {\bibinfo
  {journal} {Phys. Rev. X}\ }\textbf {\bibinfo {volume} {13}},\ \bibinfo
  {pages} {021041} (\bibinfo {year} {2023})}\BibitemShut {NoStop}%
\bibitem [{\citenamefont {Polettini}\ and\ \citenamefont
  {Esposito}(2014)}]{Polettini2014}%
  \BibitemOpen
  \bibfield  {author} {\bibinfo {author} {\bibfnamefont {M.}~\bibnamefont
  {Polettini}}\ and\ \bibinfo {author} {\bibfnamefont {M.}~\bibnamefont
  {Esposito}},\ }\href {\doibase 10.1063/1.4886396} {\bibfield  {journal}
  {\bibinfo  {journal} {J. Chem. Phys.}\ }\textbf {\bibinfo {volume} {141}},\
  \bibinfo {eid} {024117} (\bibinfo {year} {2014}),\
  10.1063/1.4886396}\BibitemShut {NoStop}%
\bibitem [{\citenamefont {Caplan}(1966)}]{Caplan1966}%
  \BibitemOpen
  \bibfield  {author} {\bibinfo {author} {\bibfnamefont {R.~S.}\ \bibnamefont
  {Caplan}},\ }\href {\doibase https://doi.org/10.1016/0022-5193(66)90124-X}
  {\bibfield  {journal} {\bibinfo  {journal} {J. Theor. Biol.}\ }\textbf
  {\bibinfo {volume} {10}},\ \bibinfo {pages} {209} (\bibinfo {year}
  {1966})}\BibitemShut {NoStop}%
\bibitem [{\citenamefont {Wachtel}\ \emph {et~al.}(2022)\citenamefont
  {Wachtel}, \citenamefont {Rao},\ and\ \citenamefont
  {Esposito}}]{Wachtel2022}%
  \BibitemOpen
  \bibfield  {author} {\bibinfo {author} {\bibfnamefont {A.}~\bibnamefont
  {Wachtel}}, \bibinfo {author} {\bibfnamefont {R.}~\bibnamefont {Rao}}, \ and\
  \bibinfo {author} {\bibfnamefont {M.}~\bibnamefont {Esposito}},\ }\href
  {\doibase 10.1063/5.0091035} {\bibfield  {journal} {\bibinfo  {journal} {The
  Journal of Chemical Physics}\ }\textbf {\bibinfo {volume} {157}} (\bibinfo
  {year} {2022}),\ 10.1063/5.0091035}\BibitemShut {NoStop}%
\bibitem [{\citenamefont {Bilancioni}\ and\ \citenamefont
  {Esposito}(2025)}]{Bilancioni2025}%
  \BibitemOpen
  \bibfield  {author} {\bibinfo {author} {\bibfnamefont {M.}~\bibnamefont
  {Bilancioni}}\ and\ \bibinfo {author} {\bibfnamefont {M.}~\bibnamefont
  {Esposito}},\ }\href {https://arxiv.org/abs/2405.17960} {\bibfield  {journal}
  {\bibinfo  {journal} {Nature communication}\ } (\bibinfo {year} {2025})},\
  \Eprint {http://arxiv.org/abs/2405.17960} {arXiv:2405.17960 [q-bio.MN]}
  \BibitemShut {NoStop}%
\bibitem [{\citenamefont {Vroylandt}\ \emph {et~al.}(2018)\citenamefont
  {Vroylandt}, \citenamefont {Lacoste},\ and\ \citenamefont
  {Verley}}]{Vroylandt2018}%
  \BibitemOpen
  \bibfield  {author} {\bibinfo {author} {\bibfnamefont {H.}~\bibnamefont
  {Vroylandt}}, \bibinfo {author} {\bibfnamefont {D.}~\bibnamefont {Lacoste}},
  \ and\ \bibinfo {author} {\bibfnamefont {G.}~\bibnamefont {Verley}},\ }\href
  {\doibase 10.1088/1742-5468/aaa8fe} {\bibfield  {journal} {\bibinfo
  {journal} {Journal of Statistical Mechanics: Theory and Experiment}\ }\textbf
  {\bibinfo {volume} {2018}},\ \bibinfo {pages} {023205} (\bibinfo {year}
  {2018})}\BibitemShut {NoStop}%
\bibitem [{\citenamefont {Raux}\ \emph {et~al.}(2024)\citenamefont {Raux},
  \citenamefont {Goupil},\ and\ \citenamefont {Verley}}]{Raux2024}%
  \BibitemOpen
  \bibfield  {author} {\bibinfo {author} {\bibfnamefont {P.}~\bibnamefont
  {Raux}}, \bibinfo {author} {\bibfnamefont {C.}~\bibnamefont {Goupil}}, \ and\
  \bibinfo {author} {\bibfnamefont {G.}~\bibnamefont {Verley}},\ }\href
  {\doibase 10.1103/physreve.110.014134} {\bibfield  {journal} {\bibinfo
  {journal} {Physical Review E}\ }\textbf {\bibinfo {volume} {110}},\ \bibinfo
  {pages} {014134} (\bibinfo {year} {2024})}\BibitemShut {NoStop}%
\bibitem [{\citenamefont {Raux}\ \emph {et~al.}(2025)\citenamefont {Raux},
  \citenamefont {Goupil},\ and\ \citenamefont {Verley}}]{Raux2025thermoelec}%
  \BibitemOpen
  \bibfield  {author} {\bibinfo {author} {\bibfnamefont {P.}~\bibnamefont
  {Raux}}, \bibinfo {author} {\bibfnamefont {C.}~\bibnamefont {Goupil}}, \ and\
  \bibinfo {author} {\bibfnamefont {G.}~\bibnamefont {Verley}},\ }\href
  {https://arxiv.org/abs/2412.15036} {\  (\bibinfo {year} {2025})},\ \Eprint
  {http://arxiv.org/abs/2412.15036} {arXiv:2412.15036 [cond-mat.stat-mech]}
  \BibitemShut {NoStop}%
\bibitem [{\citenamefont {Raux}(2024)}]{raux:tel-04964379}%
  \BibitemOpen
  \bibfield  {author} {\bibinfo {author} {\bibfnamefont {P.}~\bibnamefont
  {Raux}},\ }\emph {\bibinfo {title} {{Circuit theory for thermodynamic engines
  in stationary nonequilibrium}}},\ \href
  {https://theses.hal.science/tel-04964379} {\bibinfo {type} {Theses}},\
  \bibinfo  {school} {{Universit{\'e} Paris Cit{\'e}}} (\bibinfo {year}
  {2024})\BibitemShut {NoStop}%
\end{thebibliography}%
\end{document}